\documentclass{revtex4}
\usepackage{graphicx}
\begin{document}

\setlength{\textwidth}{16.5cm}
\setlength{\textheight}{22.2cm}
\setlength{\voffset}{1.3cm}

\newcommand{\tautau}{\ensuremath{\tau^+\tau^-}}
\newcommand{\babar}{BaBar}
\newcommand{\fbi}{fb$^{-1}$}
\newcommand{\fiveprong}{\ensuremath{\tau^-\to 3h^-2h^+\nu_{\tau}}}
\newcommand{\fone}{\ensuremath{\tau^-\to f_1(1285) \pi^- \nu_{\tau}}}
\newcommand{\sevenprong}{\ensuremath{\tau^-\to 4\pi^-3\pi^+(\pi^0)\nu_{\tau}}}
\newcommand{\BR}{\ensuremath{ B}}
\newcommand{\sevenprongnopizero}{\ensuremath{\tau^-\to 4\pi^-3\pi^+ \nu_{\tau}}}
\newcommand{\sevenprongonepizero}{\ensuremath{\tau^-\to 4\pi^-3\pi^+ \pi^0\nu_{\tau}}}
\newcommand{\taupipi}{\ensuremath{\tau\to\pi\pi^0\nu}}

\title{$\tau$ PHYSICS AT $B$-FACTORIES}

\author{Olga Igonkina \\
        {\em Physics Department, University of Oregon,
Eugene, Oregon 97403, USA}\\
}

\baselineskip=11.6pt
\begin{abstract}
Today the $B$-factories \babar{} and Belle have accumulated largest samples
of \tautau{} events and are competing to be called
$\tau$-factories. Among the problems to be tested and measurements to
be done by \babar{} and Belle are check of CP and CPT invariance in
tau decays, measurement of strange and non-strange spectral functions,
extraction of mass of strange quark and $|V_{us}|$, searches for lepton
flavor violation processes. In this paper, the latest results in tau
physics by these two experiments and 
measurements to be done within next few years are reviewed.
\end{abstract}
\maketitle
\baselineskip=14pt
\section{Introduction}

\babar{}\cite{Babar_detector} and Belle\cite{Belle_detector} are
the $e^+e^-$ collider experiments running at $\sqrt{s}$ equal to
$\Upsilon(4S)$ mass. In spite of their commonly used name {\it
$B$-factories} they also provide largest and cleanest samples today 
for study of tau physics. Indeed, the cross-section of $e^+e^-\to
\tautau$ ($\sigma_{\tautau}$) at this energy is 0.89~nb, of the same
order as $B\bar{B}$ cross section at $\Upsilon(4S)$ and just a bit
smaller than $\sigma_{\tautau}^{thr} \simeq 1.2$~nb at \tautau{} production
threshold\cite{tauAtThreshold}. 

The \tautau{} event produced at $\Upsilon(4S)$ has very characteristic
topology. The decay products of two taus are well separated in space,
such that if space is split on two hemispheres with respect to the
axis of the event thrust, the decay products of two taus are mostly
contained within opposite hemispheres. From other side, the boost is
not as large as at LEP experiments, and the tracks are well separated. As
both \babar{} and Belle are multipurpose spectrometers, the full
particle identification of the event can be performed. Usually tau
decays have one (1-prong) or three (3-prong) charged particles in the
final state, therefore the multiplicity of the \tautau{} events is
relatively small. The typical backgrounds are radiative Bhabha and
di-muon events which can be suppressed by vetoing leptons or high
momentum tracks, and hadronic $q\bar{q}$ events which are more
isotropic and have in average more neutral particles.

Currently, \babar{} has recorded about $6 \cdot 10^{8}$ tau decays,
and Belle has recorded about $10^{9}$ tau decays with large part of
the statistics used in the analyzes presented below. In
section~\ref{sec:SM}, we review high precision measurements of tau
mass and lifetime, which allow test of CPT invariance, and discuss
perspectives of measuring CP-violation in tau
decays. Section~\ref{sec:SF} is concentrated on the description of the
measurements of the hadronic tau decays. In the section~\ref{sec:LFV}
the searches of lepton flavor violation in tau decays are described
and the conclusions are drawn in section~\ref{sec:conclusions}. The
future perspectives and limitations of different measurements are
discussed throughout the paper.

\section{Standard Model tests}
\label{sec:SM}

The basic tasks of the $\tau$-factories are to measure mass $m_{\tau}$
and time of life $\tau_{\tau}$ of tau lepton. 
Belle has recently presented a preliminary measurement of $m_{\tau} =
(1776.71 \pm 0.25_{stat} \pm 0.62_{sys})$~MeV\cite{taumassBelle}
using pseudo-mass technique pioneered by ARGUS\cite{taumassARGUS}. The
sample of 253~\fbi{} was used. Although the measurement is dominated by
systematic uncertainties, it is largely due to the size of the control
samples, and therefore is likely to be improved with increased
statistics. The sample is also used to probe the difference between
$m_{\tau^+}$ and $m_{\tau^-}$ which is found to be negligible,
$|m_{\tau^+} - m_{\tau^-}|/m_{\tau} < 5.0 \cdot 10^{-4}$ at 90\%
confidence level (CL). This number is statistically limited as most
systematic uncertainties are canceled in the ratio.

At the same time \babar{} has concentrated on the tau lifetime
measurement. The flight distance transverse to the beam $\lambda_{T}$
is measured and corrected by polar angle of 3-prong system
($\Theta_{3pr}$) to calculate the total decay length $\lambda =
\lambda_T /\sin{\Theta_{3pr}}$. The dependence on azimuthal angle
$\phi$ ($\lambda(\phi)$) is fitted to minimize the systematic
uncertainties due to alignment of the vertex detector. The preliminary
result is $\tau_{\tau} = 289.40 \pm 0.91_{stat} \pm 0.90_{sys}$~fs\cite{taulivetimeBabar}. It is in agreement with PDG value
$\tau_{\tau} = 290.6 \pm 1.1$~fs\cite{PDG} and is the most precise measurement
up to date. 80~\fbi{} were used in the analysis. As in case of
$m_{\tau}$ measurement, the systematic uncertainty are partially
limited by statistics of control samples and is likely to be improved
with luminosity, although not more than by factor of two.  The
preliminary study of $\tau_{\tau^+}$ and $\tau_{\tau^-}$ showed no difference
${\tau_{\tau^-} - \tau_{\tau^+}}/{\tau_{\tau^-} + \tau_{\tau^+}} =
(0.12 \pm 0.32_{stat})\%$, where the systematic uncertainty is to be
estimated but likely to be small.

Using the above numbers averaged with PDG values and leptonic branching fractions\cite{PDG} of tau 
one can compare lepton charged current coupling constants:
\begin{equation}
\frac{g_e}{g_{\mu}} = \sqrt{ \frac{\BR(\tau\to e\nu\nu)}{\BR(\tau\to\mu\nu\nu)}
\frac{(1+C_{\tau \mu})}{(1+C_{\tau e})} }= 0.9997 \pm 0.0024
\end{equation}
\begin{equation}
\frac{g_{\mu}}{g_{\tau}} = \sqrt{
\frac{(1+C_{\tau e})}{(1+C_{\mu e})}
\frac{\tau_{\tau}}{\tau_{\mu}}
(\frac{m_{\tau}}{m_{\mu}})^5 \frac{1}{\BR(\tau\to e\nu\nu)} }= 0.9980 \pm 0.0022,
\end{equation}
where $C_{\tau e} = -0.004$, $C_{\tau \mu} = -0.0313$ and $C_{\mu e} =
-0.0044 $ are radiative corrections. No significant deviation from SM
is observed.

The subject which is still in {\it to do} list of both experiments is
a search of $CP$-violation in tau decays. While no such $CP$-violation
is expected in SM, other contribution, like e.g. 
charged Higgs exchange can result in non-negligible effect in angular
and visible mass distributions of tau decay products due to
interference of vector and scalar parts. CLEO has searched for such
effect in the decays $\tau\to K_S^0 \pi\nu$\cite{KcpCLEO} and
\taupipi{}\cite{picpCLEO} with 13.3~\fbi. While no signal was found,
the CLEO collaboration has put limits on imaginary part of charged Higgs coupling
 of $-0.172< Im(\Lambda) < 0.067 $ from
$\tau^-\to K^0_S \pi^-\nu$ data assuming $K^*(1430)$ scalar
contribution and $-0.046 <Im(\Lambda) < 0.022$ from
\taupipi{} for maximal scalar contribution. The limits are at 90\%
CL. The largest source of the uncertainty here is the size of the
sample recorded. Given that $B$-factories have almost two orders of
magnitude more data, it should be possible to improve CLEO result
significantly. It is clear, however, that the understanding of the
systematic uncertainty will require a careful work, in particular,
study of possible charge asymmetry in the detector.

Of course, there are more SM tests to be performed, such as
measurement of tau electric and anomalous magnetic dipole moments,
Michael parameters, measurement of $\nu_{\tau}$ helicity. However, it
is unlikely, that either \babar{} or Belle will be able to improve
previous measurements soon.

\section{Study of hadronic tau decays}
\label{sec:SF}
Due to simplicity of the SM tau decays involving $W^-$ exchange, it is
possible to study the hadronization process in details. All hadronic
tau decays are of interest, starting from the most common 1-prong
\taupipi{} up to not yet observed 7-prong tau decay. The
analysis of spectral function of $\pi\pi^0$ is to be used for
comparing the measurement of anomalous magnetic moment of muon with SM
prediction. The analysis of tau strange decays provides an
information on mass of strange quark and $|V_{us}|$ element of CKM
matrix. The 5-prong tau decays are  studied with large statistics
and not observed 7-prong decays are used to probe non-SM
contributions.

\subsection{Non-strange spectral function} 

\begin{figure}[t]
\includegraphics[width=0.4\textwidth, height=4.5cm]{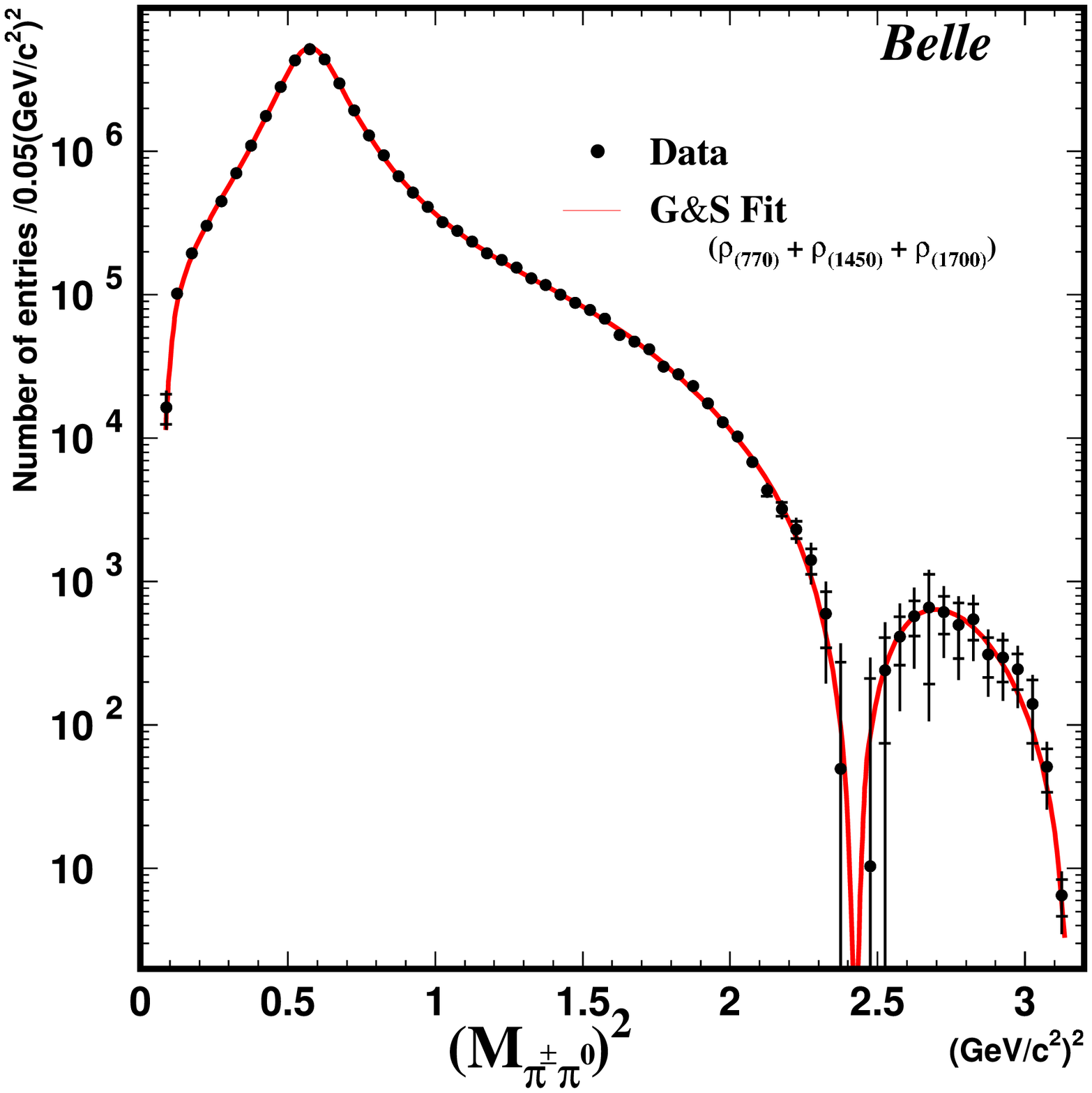}
\includegraphics[width=0.59\textwidth, height=4.2cm]{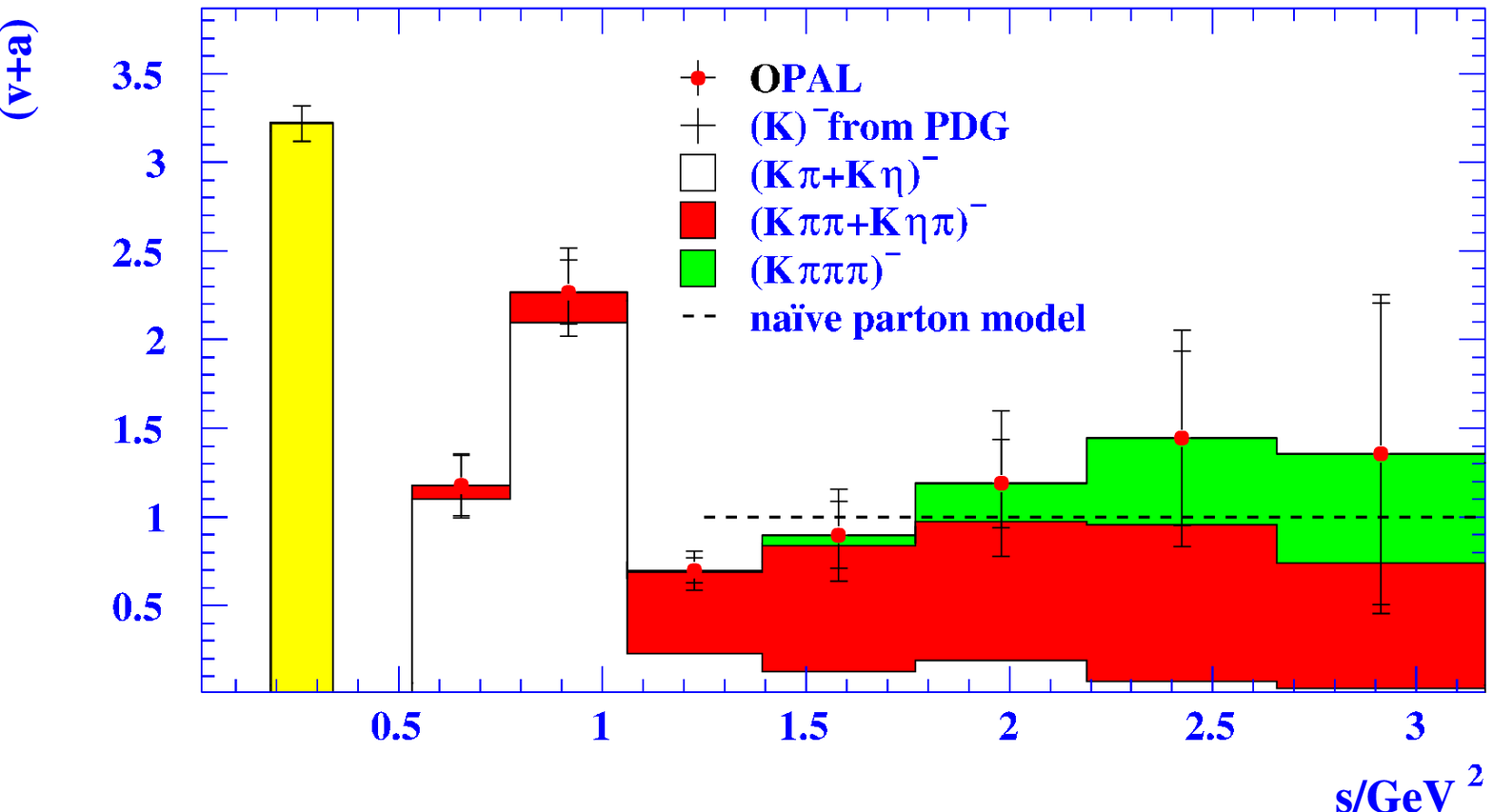}
\put(-405,50){$a)$}
\put(-45,100){$b)$}
 \caption{\it
     a$)$ Fully corrected $m_{\pi\pi^0}^2$ distribution for \taupipi. 
	The solid curve is the result of a fit to the Gounaris-Sakurai model with 
	$\rho(770)$, $\rho'(1450)$ and $\rho''(1770)$ resonances. 
     b$)$ The spectral function from strange tau decays. The dots show the inclusive 
    spectrum as measured in OPAL\cite{VusOPAL}. The histograms show exclusive spectra 
    as described on the plot. 
    \label{fig:hadrondecay} }
\end{figure}

The calculation of hadronic part of the anomalous magnetic moment of muon
$a_{\mu}^{had,LO}$ includes integral of the cross-section $e^+e^- \to
$~hadrons multiplied with QED kernel $K(s)$. The structure of $K(s)$
is such, that 75\% of $a_{\mu}^{had,LO}$ is covered by two pion final
state dominated by $\rho(770)$ resonance. Assuming isospin invariance,
$\sigma(e^+e^-\to \pi^+\pi^-)$ can be estimated from the branching
fraction $\BR(\tau^-\to \pi^-\pi^0\nu)$\cite{amutheory}. Currently,
the results based on tau data together with results of Muon g-2
experiment\cite{g-2} give $a_{\mu}^{exp} - a_{\mu}^{SM} = (9.4 \pm
10.5) \cdot 10^{-10}$, while calculation based solely on
$e^+e^-$ data is $a_{\mu}^{exp} - a_{\mu}^{SM} = (25.2 \pm 9.2) \cdot
10^{-10}$\cite{amutheory}.
Belle has recently presented new preliminary study of \taupipi{}
decay\cite{rhoBelle}. The measured $\pi^-\pi^0$ invariant mass spectrum is corrected
for the detector deficiency and distortions using the unfolding technique
and then fitted with Gounaris-Sakurai function  as
shown on Fig.~\ref{fig:hadrondecay}a. The distribution also exposes
$\rho''$ resonance, evident in this decay for the first time. The
obtained $\pi\pi$ contribution to $a_{\mu}$ is $a_{\mu}^{\pi\pi} =
(462.4 \pm 0.6_{stat} \pm 3.2_{sys} \pm 2.3_{isospin}) \cdot
10^{-10}$, which yields $a_{\mu}^{exp} - a_{\mu}^{SM} = (11.0 \pm 10.5)
\cdot 10^{-10}$, in good agreement with ALEPH and CLEO data. Although,
only 72~\fbi{} were used in this analysis, the systematic uncertainties,
such as on the track and $\pi^0$ reconstruction efficiency, dominate
and it is unlikely to be improved with larger statistics.

\subsection{Strange spectral function}

Analysis of strange tau decays allows to extract mass of strange quark
$m_s$ and $|V_{us}|$ element of CKM matrix  via moments of the strange spectral function (SSF):
\begin{equation}
R^{kl}_{\tau} = \int_0^{M_{\tau}^2} ds (1-\frac{s}{m_{\tau}^2})^k (\frac{s}{m_{\tau}^2})^l
\frac{\BR(\tau\to X^{(S=-1)}\nu)}{\BR(\tau\to e\nu\nu)} \frac{dN_{X^{(S=-1)}}}{N_{X^{(S=-1)}}ds}. 
\end{equation}
$R_{\tau}^{kl}$ are calculable within operator product expansion
framework with phenomenological hadronic parametrization\cite{Vustheory}.

The $R_{\tau}^{(0,0)}$ moment is most sensitive to the $|V_{us}|$,
while its dependence on $m_s$ is small and can be
neglected\cite{Vustheory}. This allows to extract $|V_{us}| = 0.2208
\pm 0.0033_{exp} \pm 0.0009_{th}$ from results of OPAL\cite{VusOPAL}
assuming $m_s(2\rm{GeV}) = 95$~MeV. The value is already very
competitive with the estimate from $K\to\pi e \nu$ decays $|V_{us}| =
0.2200 \pm 0.0026$\cite{PDG} and unlike in $K\to\pi e \nu$ case the
theoretical uncertainty is significantly smaller than
experimental. Higher order moments are more sensitive to the $m_s$ and
one extracts $m_s(2\rm{GeV}) = (81 \pm 22)$~MeV from the same
data. The authors of \cite{Vustheory} anticipate simultaneous
extraction of $m_s$ and $|V_{us}|$ from the data in future.

The OPAL result is largely statistics limited with total of 162
thousands of identified tau events. While there is only a
preliminary result from
\babar{} on $\BR(\tau\to K \pi^0\nu) = (4.38\pm 0.04_{stat} \pm
0.22_{sys})\cdot 10^{-3}$ available, one can expect a significant
improvement of knowledge of  SSF from $B$-factories. The
statistical uncertainties is very small, and systematic uncertainties
are largely correlated for different $\tau\to X^{(S=-1)}\nu$ exclusive
channels, and the measurement is expected to be few times more
precises than current PDG value of $\BR(\tau\to X ^{(S=-1)} \nu) =
(29.1\pm 0.8)\cdot 10^{-3}$\cite{PDG}. 

\subsection{5- and 7-prong tau decays}

\begin{figure}[t]
\includegraphics[width=0.33\textwidth, height=4.5cm]{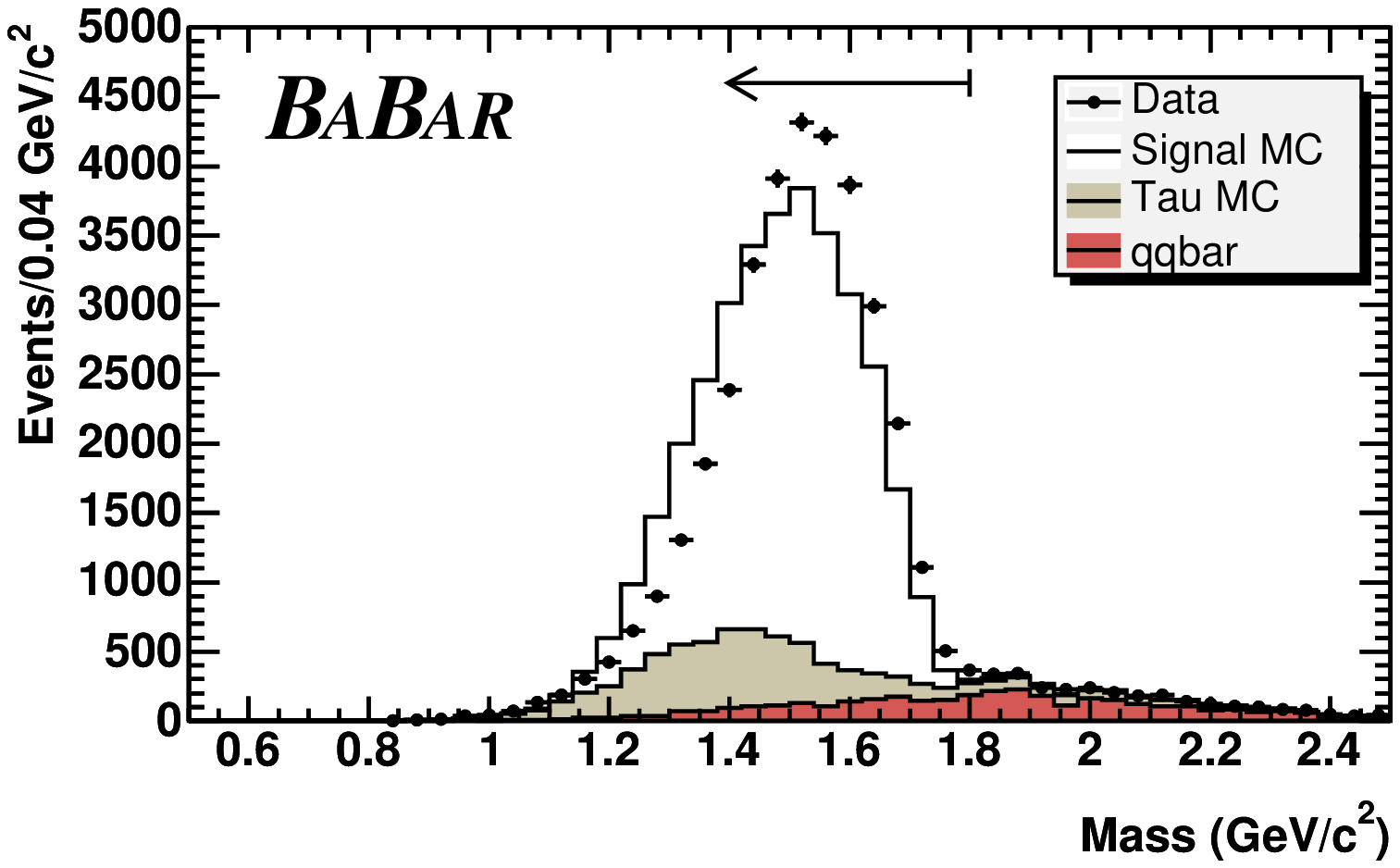}
\includegraphics[width=0.33\textwidth, height=4.5cm]{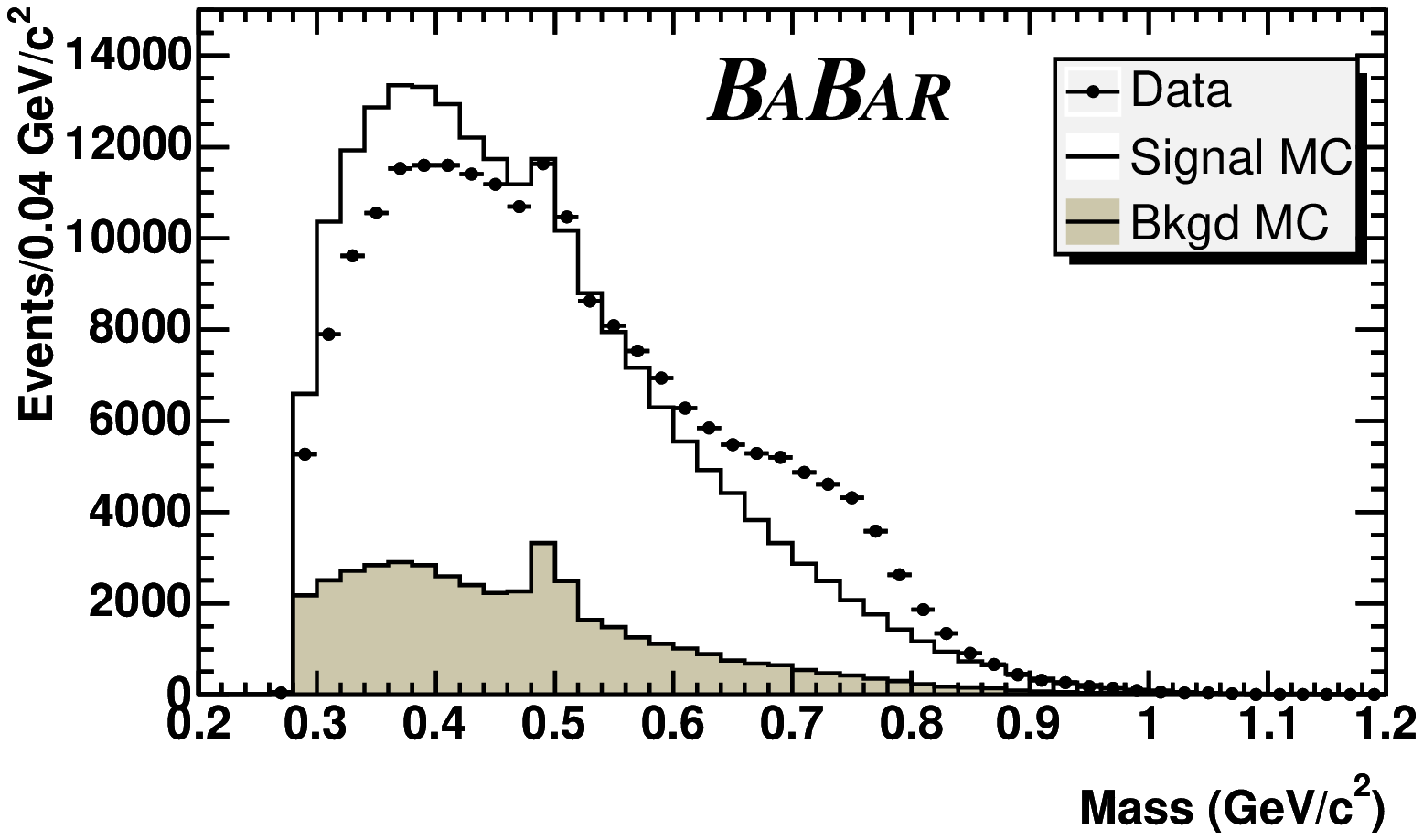}
\includegraphics[width=0.30\textwidth, height=4.5cm]{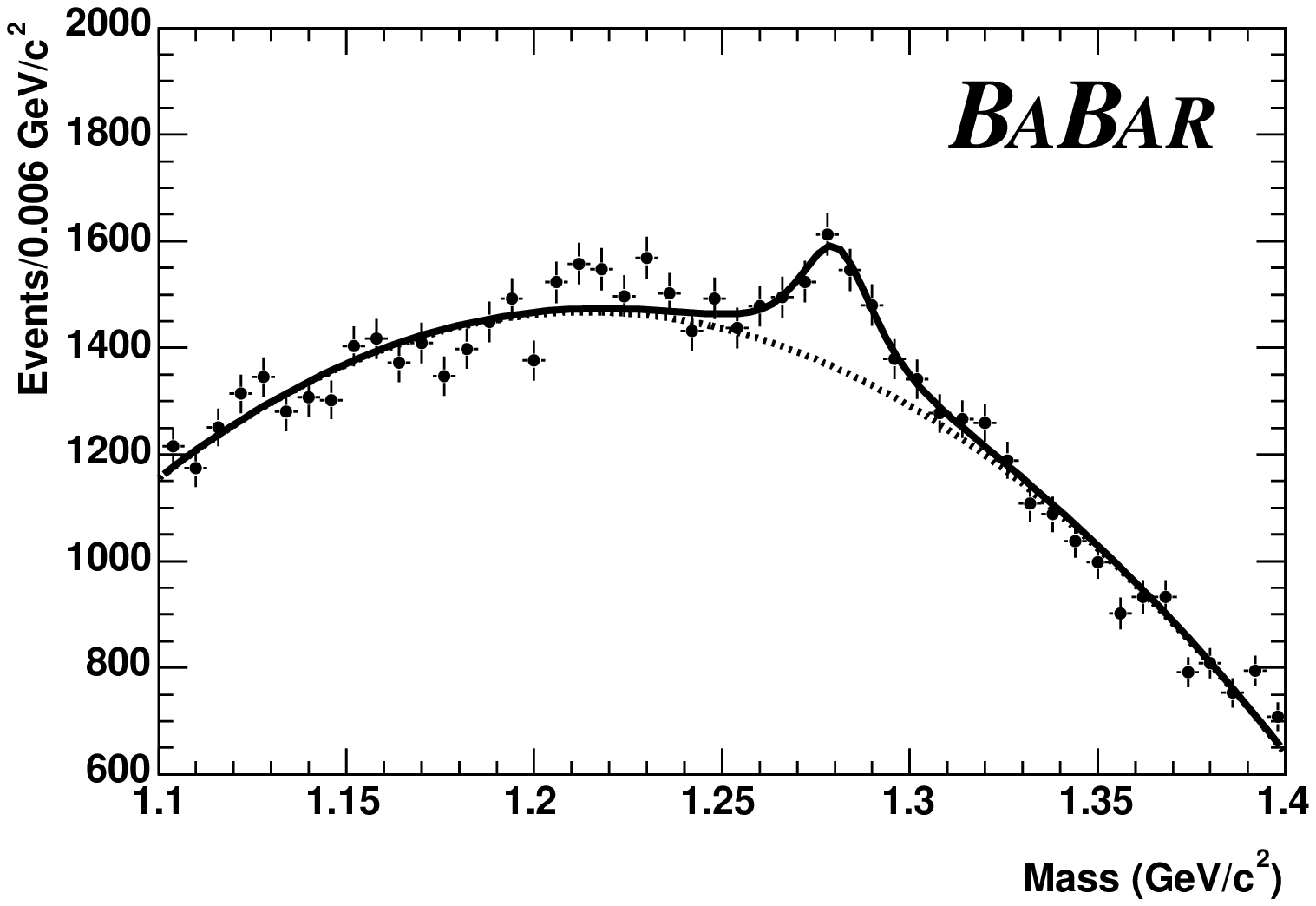}
\put(-410,45){$a)$}
\put(-185,45){$b)$}
\put(-80,45){$c)$}
\caption{\it Invariant mass of $a)$ five charged particles
$b)$ $h^+h^-$ pairs $c)$ $2h^+2h^-$ combinations. All tracks are taken
as pions. The points are the data and the histograms are the Monte
Carlo simulation. The unshaded and shaded histograms are the signal
and background events, respectively. The Monte Carlo sample is
normalized to the luminosity of the data sample.  The solid line on
plot $c$ is a fit to the data. \label{fig:5prong} }
\end{figure}

With such big sample at hand, one can look into underlying structure
of rare tau decays. \babar{} has recently published a study of 5-prong
decays $\tau^- \to 3h^- 2h^+ \nu$\cite{5prongBabar}. The branching
fraction is $(8.56 \pm 0.05 \pm 0.42)\cdot 10^{-4}$ in agreement with
previous measurements\cite{PDG}.  However, the invariant mass of five
hadrons is different from the phase-space distribution assumed before
(see Fig.~\ref{fig:5prong}). The contribution of $\rho$ meson is
evident in the mass of two pion, and $f_1$ resonance is observed in
the four pion mass distribution, $\BR(\tau^-\to f_1 h^- \nu_{\tau}) =
(3.9\pm 0.7\pm 0.5)\cdot 10^{-4}$\cite{5prongBabar}.
\babar{} has also searched for 7-prong tau decays. If observed, they
would signal of non-SM contribution, as SM predicts $\BR(\tau^-\to
4\pi^-3\pi^+\nu) < 10^{-9}$. No signal is found in either exclusive or
inclusive 7-prong tau decays and the obtained upper limits are
\begin{equation}
\begin{array}{lcl}
       \BR(\sevenprong)                             &  < &3.0 \cdot 10^{-7} \\
       \BR(\tau^- \to 4\pi^-3\pi^+ \nu_{\tau})      &  < &4.3 \cdot 10^{-7} \\
       \BR(\tau^- \to 4\pi^-3\pi^+ \pi^0 \nu_{\tau})&  < &2.5 \cdot 10^{-7} \\
\end{array}
\end{equation}
at 90\% CL\cite{7prongBabar}. 232~\fbi{} is used in both analyzes.

\section{Searches for Lepton Flavor Violation} 
\label{sec:LFV}

One of the most interesting question in tau physics now is there a
sizable lepton flavor violation (LFV) or not. Given the observation of
neutrino oscillation by experiment\cite{neutrinos}, one expects
charged LFV even in Standard Model extended with massive
neutrinos. However, the expected branching fractions are negligible
and far beyond reach of the current experiments. From other side, many
other extensions of SM, e.g. supersymmetry, predict LFV on the level
of $10^{-10} - 10^{-7}$\cite{LFVtheory}, which can be probed with
currently accumulated statistics. Analysis of different channels is
important. While $\tau\to\mu\gamma$ is expected to be the largest
tau LFV decay in most models,  $\tau\to 3 \ell$ can expose
supersymmetric Higgs contribution\cite{LFVHiggs}, and $\tau^-\to
\ell^+ h^-h^-$ violates not only lepton flavor, but also lepton
number. If LFV process would be observed in an experiment, the combined
analysis of different channels will allow to understand underlying mechanism and to
differentiate between the models. 

Both \babar{} and Belle are very active in searches of lepton flavor
violation. Unfortunately, no signal is found in any channel and upper
limits are set on the level of $10^{-7}$ (see table~\ref{tab:LFV}).
Both experiments plan to increase their samples by factor of 2 by
2008. 

\begin{table}[t]
\centering
\caption{ \it 90\% CL upper limits on LFV tau decays obtained by $B$-factories\cite{LFVexp}. 
Numbers given in $10^{-7}$ units.  Second column for each experiment shows integrated luminocity used in analysis.
}
\vskip 0.1 in
\begin{tabular}{|l|c|c|c|c|} \hline
          & \multicolumn{2}{|c|}{ \babar{}} &\multicolumn{2}{|c|}{ Belle }\\
\hline
\hline
 Channel                                & UL    & ${\cal L}$    & UL    & ${\cal L}$\\ 
\hline
$\tau^-\to\mu^-\gamma$          	& 0.7   & 232~\fbi{} 	& 3.1 & 86~\fbi{} \\
$\tau^-\to e^-\gamma$          		& 1.1   & 232~\fbi{} 	& 3.9 & 87~\fbi{} \\
\hline
$\tau^-\to e^-e^+e^-$                   & 2.0   & 91~\fbi{}	& 3.5 & 87~\fbi{} \\
$\tau^-\to\mu^-\mu^+\mu^-$              & 1.9   & 91~\fbi{}	& 2.0  & 87~\fbi{} \\
$\tau^-\to \ell^-\ell^{\pm}\ell^{'\mp}$ & (1-3) & 91~\fbi{}	& (2-4) & 87~\fbi{} \\
\hline 
$\tau^-\to \ell^- h^+h^-$       	& (1-3)  & 221~\fbi{} 	& &  \\
$\tau^-\to \ell^+ h^-h^-$       	& (0.7-5)& 221~\fbi{} 	& &  \\
\hline
$\tau^-\to \ell^- \pi^0,\eta,\eta'$     &        &            	& 2-10& 154~\fbi{}\\
\hline
$\tau^-\to\Lambda\pi^-$         	&        &		& 0.7 & 154~\fbi{}\\
$\tau^-\to{\bar{\Lambda}}\pi^-$  	&        &		& 1.4 & 154~\fbi{}\\

\hline
\end{tabular}
\label{tab:LFV}
\end{table}

\begin{figure}[t]
\includegraphics[width=0.48\textwidth, height=4.5cm]{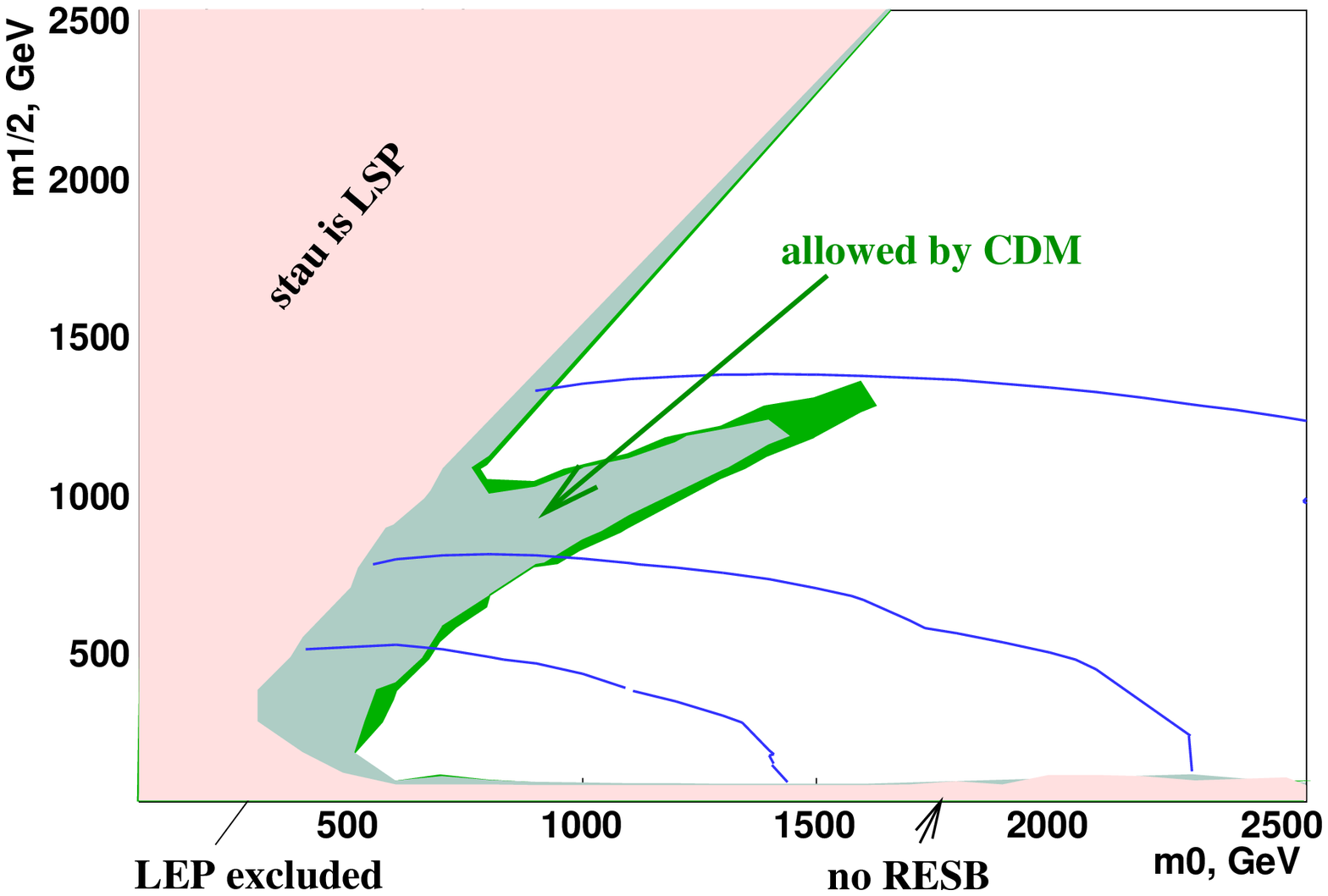}
\includegraphics[width=0.48\textwidth, height=4.5cm]{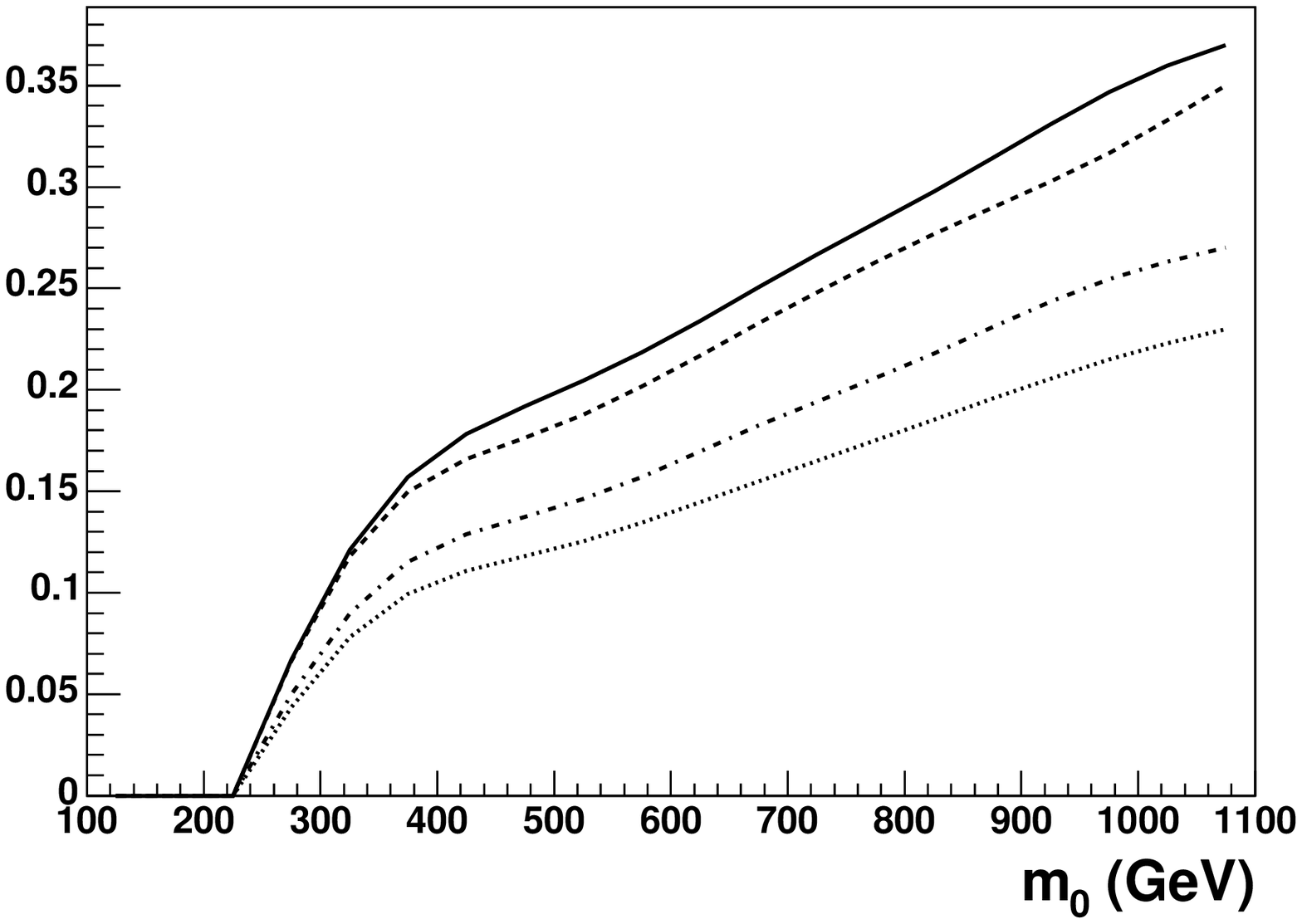}
\put(-305,100){\bf{$a)$}}
\put(-135,100){\bf{$b)$}}
\put(-170,50){\rotatebox{90}{$M^2_{L23}/M^2_{L22}$}}
 \caption{\it
     a$)$ Exclusion plot of mSUGRA with right handed neutrinos as function of
	gaugino ($m_{1/2}$) and scalar ($m_0$) masses for $\tan{\beta} = 50$.
	The green area is allowed by cold dark matter searches and blue curves show
        area excluded by $\tau\to\mu\gamma$ as described in the text.
     b$)$ Model independent upper limits of off-diagonal element of slepton mixing matrix
     $M^2_{L23}/M^2_{L22}$ as function of $m_0$ from $\BR(\tau\to\mu\gamma)$ as described in the text.
    \label{fig:LFV} }
\end{figure}

The obtained limits can already be used to restrict parameter space of
the models. The Fig~\ref{fig:LFV}a\cite{igonSUSY05} shows the
exclusion plot for mSUGRA with right handed neutrinos as function of
gaugino ($m_{1/2}$) and scalar ($m_0$) masses at grand unification scale
$m_{GUT} = 5\cdot 10^{15}$~GeV for $\tan{\beta} = 50$. The latest
measurements of neutrino mixing matrix and masses\cite{neutrinoExp}
are used for Yukawa couplings.  The mass of right handed neutrinos is
set to $M_{\nu_R} = 5 \cdot 10^{14}$ and normal hierarchy is assumed
for left handed neutrinos. Everything but a green area is excluded by
theory or cold dark matter density measurement, while the area below blue curves
(towards center of coordinates) is excluded by tau LFV searches. The
blue curves from center of coordinates onwards correspond to the
$6.8\cdot 10^{-8}$, $1\cdot 10^{-8}$ and $1\cdot 10^{-9}$. One can
expect that by 2008 the combined sensitivity of \babar{} and Belle
will reach $1\cdot 10^{-8}$ level, while running super-$B$ factory
will be necessary to reach $1\cdot 10^{-9}$ level.

Figure~\ref{fig:LFV}b\cite{igonSUSY05} shows the upper limits set on
off-diagonal elements of slepton mixing matrix $M^2_{L23}/M^2_{L22}$
({\it model independent approach}) as function of $m_0$ for
$\tan{\beta} = 50$ and $m_{1/2} = 100+0.8\cdot m_0$. The uppermost
curve corresponds to $6.8\cdot 10^{-8}$, followed by curves of $5\cdot
10^{-8}$, $2\cdot 10^{-8}$ and $1\cdot 10^{-8}$.

\section{Conclusions}
\label{sec:conclusions}

Current $B$-factories have a large and interesting program to study
tau physics. The accumulated statistics reaches $10^{9}$ tau decays
which allows very precise measurements and searches of very rare or
forbidden tau decays. Among the most important measurements are mass
and tau lifetime (systematics limited) and tests of CPT/CP violation
(statistics limited). The study of the spectral function of \taupipi{}
decay would help to clarify the comparison of measurement of anomalous
magnetic moment of muon with its SM prediction, while strange
spectral functions are to be used for $m_s$ and $|V_{us}|$ estimates.

Among very interesting studies are searches of lepton flavor violation
tau decays. While \babar{} and Belle can be lucky to discover LFV and
this way to probe the physics beyond Standard Model, one would need
statistics of super-$B$ factory to be able to measure mixing of
sleptons accurately.

%

\end{document}